# An Efficient Hardware Implementation of Elliptic Curve Point Multiplication over GF ($2^m$) on FPGA

Ruby Kumari[1, 2] [0000-0003-4739-2198], Tapas Rout[2] [0009-0005-8331-7218], Babul Saini[2] [0009-0003-4834-7441], Jai Gopal Pandey[1, 2] [0000-0001-9937-7438] and Abhijit Karmakar[1, 2] [0000-0002-4681-1998]

[1]Academy of Scientific & Innovative Research (AcSIR), CSIR-CEERI Campus
[2] CSIR – Central Electronics Engineering Research Institute, Pilani, India
ruby.ceeri20a@acsir.res.in, tapasrout.riscv@outlook.com, babulsaini42@gmail.com, jai@ceeri.res.in, abhijit@ceeri.res.in

**Abstract.** Elliptic Curve Cryptography (ECC) is widely accepted for ensuring secure data exchange between resource-limited IoT devices. The National Institute of Standards and Technology (NIST) recommended implementation, such as B-163, is particularly well-suited for Internet of Things (IoT) applications. Here, Elliptic Curve Point Multiplication (ECPM) is the most time-critical and resource-intensive operation due to the finite field multiplier. This paper proposes a new implementation method of finite field multiplication using a hybrid Karatsuba multiplier, which achieves a significant improvement in computation time while maintaining a reasonable area footprint. The proposed multiplier, along with a finite field adder, squarer, and extended Euclidean inversion circuit, is used to implement an architecture for ECPM using the Montgomery algorithm. The architecture is evaluated for *GF* ($2^{163}$) on the Xilinx Virtex-7, FPGA platform, achieving a maximum frequency of 213 MHz and occupying 14195 Look-up Tables (LUTs). The results demonstrate a significant speedup in computation time and overall performance compared to other reported designs.

## 1 Introduction

In present-day electronic systems cryptography has become vital to data communication for applications such as smartphones, web-based banking, personal digital assistants, and smart cards. It ensures the confidentiality, integrity, and authenticity of data, protects against unauthorized access and tempering, and enables the establishment of secure communication channels and shared secret keys [1], [2]. Recently, Elliptic Curve Cryptography (ECC) has proved to be a powerful cryptography approach that generates security between key pairs for public key encryption by using the mathematics of elliptic curves. It is an asymmetric cryptographic system that provides equivalent security to the well-known RSA cipher [3]. In the current era, wireless networks facilitate communication between billions of devices. However, the open and unsecured nature of the internet architecture poses a significant risk of eavesdropping on private and confidential information.

ECC is gaining popularity as it provides similar security to conventional RSA, with smaller key lengths [4], [5]. For example, 163-bit ECC is similar to 1024-bit RSA [6], [7], [8]. The ECC algorithm can be modelled in C language as in [9]. The features of ECC algorithm make it suitable for applications in resource-constrained environments. The use of Field-Programmable Gate Array (FPGA) [5] technology for hardware implementation of ECC is justified by the performance and cost efficiency of today's FPGA devices [5], as well as the ability of FPGA devices to easily update the cryptographic algorithms [8]. The compute-intensive components are identified for their hardware implementations by using residue number system (RNS) and projective coordinates.

In elliptic curve cryptosystems, the fundamental operation is point multiplication $Q = kP,$ which resembles the multiplication of an elliptic curve point $P$ by a scalar integer k to provide the resulting point $Q$ [10]. The point multiplication is performed by calculating a series of point additions and point doublings [1] on the finite fields closed under prime or irreducible polynomials; hence, requiring modular operations. ECC point multiplication is a complex operation involving finite field multiplication, squaring, addition, and inversion. The performance and efficiency of the system rely significantly on the characteristics of the finite field multiplier, especially for its area and delay. As a result, researchers have developed various algorithms and architectures with the goal of achieving rapid and effective finite field multiplication.

For realizing finite field multiplication, the conventional Polynomial Algorithm [11] and Karatsuba Algorithm [12] are the most commonly used. Polynomial Algorithm has area efficiency and lower critical delay, particularly for lower bit sizes. However, as the bit sizes increase, the Karatsuba Algorithm offers greater efficiency in terms of area utilization, albeit at the expense of a higher critical path delay. This trade-off poses a challenge when seeking an optimal solution for larger bit sizes.

To address this challenge, our paper introduces a hybrid version of Karatsuba multiplier that leverages small bit sized Polynomial multipliers as building blocks to construct Karatsuba multipliers with larger bit size. By striking a balance between area and delay, this hybrid multiplier architecture effectively mitigates the aforementioned trade-off and provides an improved solution for elliptic curve point multiplication, particularly for larger bit sizes.

The proposed architecture for Elliptic Curve Point Multiplication (ECPM) integrates the aforementioned hybrid Karatsuba multiplier with a finite field adder, squarer, and Extended Euclidean inversion circuit [13]. The Extended Euclidean inversion algorithm performs inversion in the least number of cycles, while the squaring circuit is based on pre-computed equations and implemented using only XOR gates [14]. This makes the squaring circuit faster and more area-efficient than implementing the squaring operation using a multiplier circuit. The proposed architecture is evaluated in terms of area and time complexities for GF $(2^{163})$ on the Xilinx Virtex-7 FPGA platform and compared against existing architectures in the literature.

## 2 Background

### 2.1 ECC Point Multiplication by Montgomery Method

Miller [15], [16], and Koblitz separately introduced elliptic curve encryption based on the discrete logarithm problem in 1985. An elliptic curve, $E$ over a binary field is defined as

$$E : y^2 + xy = x^3 + ax^2 + b \tag{1}$$

In ECC, point addition and point doubling are the two fundamental operations used to perform arithmetic on points on an elliptic curve. These operations allow for the construction of cryptographic algorithms such as key generation, key exchange, and digital signatures.

The point multiplication $Q = kP$, where the point $P = (x, y)$ with $x, y \in$ GF $(2^m)$ is defined over the elliptic curve given by Weierstrass Equation as in Eq. (1).The product $kP$ is obtained by adding $P$ to itself $k$ times and is called point multiplication, $kP = P + P + \cdots P(k \text{ times})$ [17]. Assuming that,

$$k = k_{t-1}2^{t-1} + k_{t-2}2^{t-2} + \ldots + k_1 2^1 + k_0 2^0 = s_t \tag{2}$$

Here it may be written as,

$$s_j = 2s_{j-1} + k_{t-j} \;\forall j = 1, 2, \ldots, t \tag{3}$$

The point multiplication can be efficiently computed using Montgomery algorithm, that consists of computing at each step $s_j P$ and $(s_j + 1)P$ as a function of $s_{j-1} P$ and $(s_{j-1} + 1)P$ as follows [17]:

If $k_{t-j} = 0$ then

$$s_j P = 2(s_{j-1} P) \tag{4a}$$

$$(s_j + 1)P = (2s_{j-1} + 1)P = s_{j-1}P + (s_{j-1} + 1)P \tag{4b}$$

and if $k_{t-j} = 1$ then

$$s_j P = (2s_{j-1} + 1)P = s_{j-1} P + (s_{j-1} + 1)P \tag{5a}$$

$$(s_j + 1)P = 2(s_{j-1} + 1)P \tag{5b}$$

Initially we need to define $s_0 P = \infty$ and $(s_0 + 1)P = P$, and the algorithm for computing point multiplication is given in Algorithm 1 that are based on Eqs. (4a), (4b), (5a) and (5b).

The fact that at each step the values of both $s_j P$and $s_j P + P$ are known, allows simplifying the computation over GF $(2^m)$. For simplified computation of the algorithm, the following property is used: If $A = (x_A, y_A) \neq \infty$ and $B = (x_B, y_B) \neq \infty$ are two different points on the curve and if $A \neq -B$, then the $x$-coordinates $x_{A+B}$ and $x_{A-B}$ of $A + B$ and $A - B$ are related by the relation [17]:

$$x_{A+B} = x_{A-B} + x_B(x_A + x_B)^{-1} + (x_B(x_A + x_B)^{-1})^2 \tag{6a}$$

**Algorithm 1.** *Montgomery Method* (*Q=kP*).

```
Input: [kt-1, …, k1, k0], with kt-1 = 1,
P = (xP, yP), over Binary Field GF(2m).
Output: Q = kP
A = point at infinity; B = P;
for j in 1 ... t loop
if k(t-j) = 0 then
     A = 2A, B = A+B;
else
     A = A+B, B = 2B;
end if;
end loop;
return Q = A
```

Furthermore, if $A = s_j P$ and $B = (s_j + 1)P$ for some $j$, then $A$ - $B$ = - $P$, $x_{A-B} = x_P$, and Eq. (6a) becomes [17],

$$x_{A+B} = x_P + x_B(x_A + x_B)^{-1} + (x_B(x_A + x_B)^{-1})^2 \tag{6b}$$

If $P$ is assumed to be different from $\infty$, $A = s_j P$ and $B = (s_j + 1)P$ is always different. If at some step $x_A = x_B$ then $A = -B$ and $A + B = \infty$. Thus, we obtain [17],

$$x_{A+A} = x_A^2 + \frac{b}{x_A^2} \text{ if } x_A \neq 0 \tag{7}$$

The Algorithm 1 can now be executed with the $x$-coordinates of the successive $A$ and $B$ points. The final step will compute the missing $y$-coordinate of the result. If $P = (x_P, y_P)$, where $x_P \neq 0$, $kP = (x_A, y_A)$ and $(k + 1)P = (x_B, y_B)$, then $y_A$ can be computed as,

$$y_A = x_P^{-1}(x_A + x_P)[(x_A + x_P)(x_B + x_P) + x_P^2 + y_P] + y_P \tag{8}$$

### 2.2 Realization of ECC Point Multiplication in Projective Coordinate System

In ECC, the projective coordinate system is a mathematical representation used to perform efficient arithmetic operations on elliptic curve points. It allows for faster point addition and doubling operations compared to using affine coordinates directly [17]. In the projective coordinate system, a point *P* on the elliptic curve is represented by three coordinates: (*X*, *Y*, *Z*). Here, *X* and *Y* are the projective coordinates representing the affine coordinates, i.e., *x* and *y* coordinates of the point, whereas *Z* is an additional non-zero element in the underlying field. The Algorithm 1 represents how point multiplication using projective coordinate system is performed in Montgomery algorithm.

The Montgomery method as described in Section 2.1 can be mapped into projective coordinate system. It allows us to get rid of the inversion operation at each iteration of Montgomery algorithm [17]. We use standard projective coordinates for our implementations. With the projective mapping $x_A = X_A/Z_A$ and $x_b = X_B/Z_B$, Eq. (6b) can be written as,

$$x_{A+B} = x_P + \frac{X_B Z_A}{X_A Z_B + X_B Z_A} + \left(\frac{X_B Z_A}{X_A Z_B + X_B Z_A}\right)^2 \tag{9}$$

Assuming

$$Z_{A+B} = (X_A Z_B + X_B Z_A)^2 \tag{10}$$

From (9), $X_{A+B}$ can be re-written as,

$$\begin{aligned} X_{A+B} &= x_P Z_{A+B} + X_B Z_A (X_A Z_B + X_B Z_A) + (X_B Z_A)^2 \\ &= x_P Z_{A+B} + X_A X_B Z_A Z_B \end{aligned} \tag{11}$$

Also, Eq. (7) can be written in projective coordinate system as,

$$x_{A+A} = \left(\frac{X_A}{Z_A}\right)^2 + b\left(\frac{Z_A}{X_A}\right)^2 = \frac{X_A^4 + bZ_A^4}{(X_A^2 Z_A^2)} \tag{12}$$

Further, assuming

$$Z_{A+A} = X_A^2 Z_A^2 \tag{13}$$

$X_{A+A}$ can now be expressed in projective coordinate as

$$X_{A+A} = x_{A+A} Z_{A+A} = X_A^4 + b Z_A^4 \tag{14}$$

Finally, according to Eq. (8), we can express $y_A$ as

$$\begin{aligned} y_A = {} & (x_P + X_A/Z_A)[(X_A + x_P Z_A)(X_B + x_P Z_B) \\ & + (x_P^2 + y_P) Z_A Z_B](x_P Z_A Z_B)^{-1} + y_P \end{aligned} \tag{15}$$

The Montgomery algorithm as implemented in the projective coordinate system is given in Algorithm 2. The scheduling scheme of each iteration of Step-2 is provided in Table 1. The scheduling scheme for each iteration of the Montgomery algorithm in Step-2 consists of 6 multiplications, 5 squaring and 3 additions. As we can see squaring and addition operations are computationally much cheaper than multiplication, therefore, optimizing multiplication operation is critical for high performance ECPM.

**Table 1.** Proposed Scheduling Scheme for Point Multiplication.

| Cycles | Adder | Multiplier | Squarer |
|---|---|---|---|
| 1 | - | $X_2Z_1$ | $X_2^2$ |
| 2 | - | $X_1Z_2$ | $Z_2^2$ |
| 3 | $X_1Z_2+X_2Z_1$ | $X_1Z_1X_2Z_2$ | $X_1Z_2+X_2Z_1 \rightarrow Z_1$ |
| 4 | - | $Z_1x_P$ | $Z_2^4$ |
| 5 | $Z_1x_P+X_1Z_1X_2Z_2 \rightarrow X_1$ | $bZ_2^4$ | $X_2^4$ |
| 6 | $bZ_2^4+X_2^4 \rightarrow X_2$ | $X_2^2Z_2^2 \rightarrow Z_2$ | - |

**Algorithm 2.** Montgomery Method using Projective Coordinate System ($Q$ = $kP$).

**Input:** $[k_{t-1}, \ldots, k_1, k_0]$, with $k_{t-1} = 1$, $P = (x_P, y_P)$, over Binary Field $GF(2^m)$
**Output:** $Q = kP = (x_3, y_3)$

**Step 1: Initialization of affine to projective coordinate**

$X_1 \leftarrow x_P;\ Z_1 \leftarrow 1;\ X_2 \leftarrow x_P^4 + b;\ Z_2 \leftarrow x_P^2$

**Step 2: Main loop: Projective point addition and doubling**

for i from (t − 2) downto 0
if $k_i = 1$ then
$R \leftarrow Z_1;\ Z_1 \leftarrow (X_1Z_2 + X_2Z_1)^2$
$X_1 \leftarrow x_PZ_1 + X_1X_2RZ_2$
$R \leftarrow X_2;\ X_2 \leftarrow X_2^4 + bZ_2^4;\ Z_2 \leftarrow R^2Z_2^2$
else
$R \leftarrow Z_2;\ Z_2 \leftarrow (X_1Z_2 + X_2Z_1)^2$
$X_2 \leftarrow x_PZ_2 + X_1X_2RZ_1$
$R \leftarrow X_1;\ X_1 \leftarrow X_1^4 + bZ_1^4;\ Z_1 \leftarrow R^2Z_1^2$
end if
end for

**Step 3: Post-process: Recover projective to affine coordinate**

$x_3 = X_1/Z_1$
$y_3 \leftarrow (x_P + X_1/Z_1)[(X_1 + x_PZ_1)(X_2 + x_PZ_2) + (x_P^2 + y_P)(Z_1Z_2)](x_PZ_1Z_2)^{-1} + y_P$
return $x_3, y_3$

## 3 Proposed Hybrid Karatsuba Multiplier

The Karatsuba algorithm is an efficient polynomial multiplication algorithm. This algorithm is applied to larger degree polynomials based on splitting it into a lower and an upper half, and performing the computations recursively with other requisite operations.

Two $n$-bit polynomials $a(x) = \sum_{i=0}^{n-1} a_i x^i$ and $b(x) = \sum_{i=0}^{n-1} b_i x^i$ are split into higher$(A_H, B_H)$ and lower$(A_L, B_L)$ significant halves as follows (with $n = 2m$) [12]:

$$a(x) = x^m \sum_{i=0}^{m-1} a_{m+i} x^i + \sum_{i=0}^{m-1} a_m x^i = A_H x^m + A_L \tag{16a}$$

$$b(x) = x^m \sum_{i=0}^{m-1} b_{m+i} x^i + \sum_{i=0}^{m-1} b_m x^i = B_H x^m + B_L \tag{16b}$$

Using Karatsuba Algorithm (KA),

$$a(x)b(x) = M_2(x)x^{2m} + [M_2(x) + M_1(x) + M_0(x)]x^m + M_0(x) \tag{17}$$

where, $M_2 = A_H B_H, M_1 = (A_H + A_L)(B_H + B_L), M_0 = A_L B_L$

Eq. (17) shows that three sub-multipliers are required in order to obtain the multiplication results using KA. The schematic realization of Karatsuba Algorithm is shown in Fig. 1 that comprises of two main blocks. The first one includes splitting, sub-multiplication and alignment stages, followed by an overlap circuit, i.e. the second block which adds the common powers of *x* in the generated product.

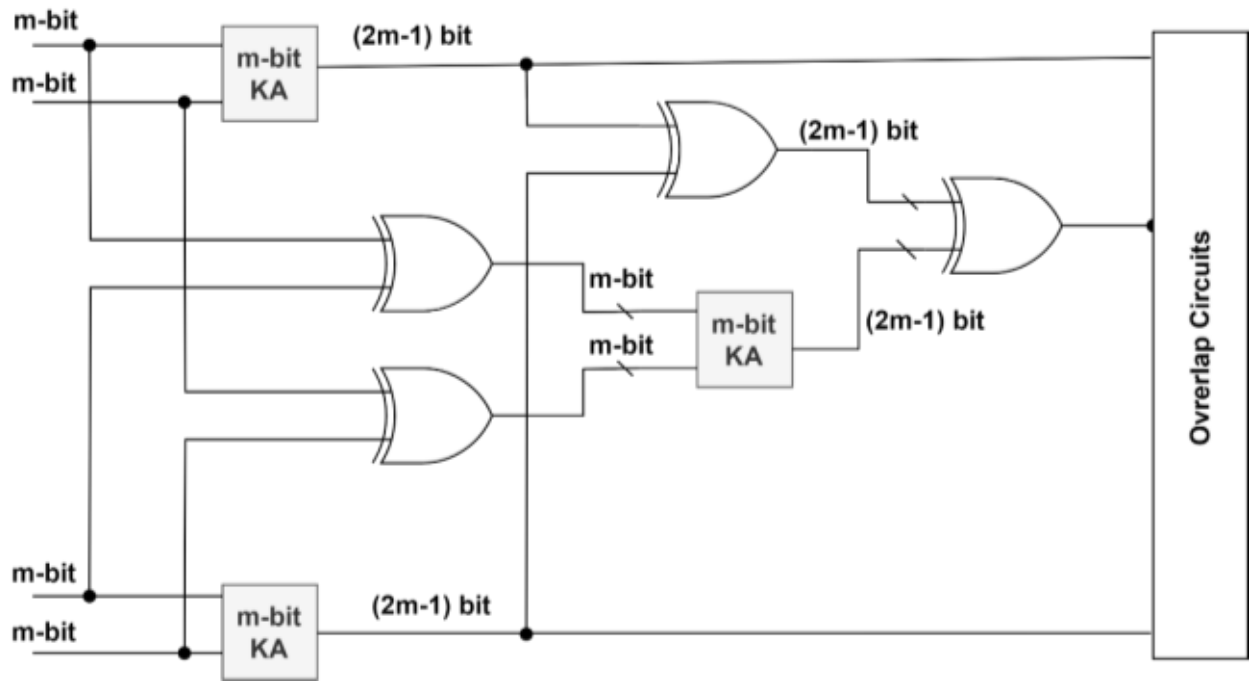

(a)

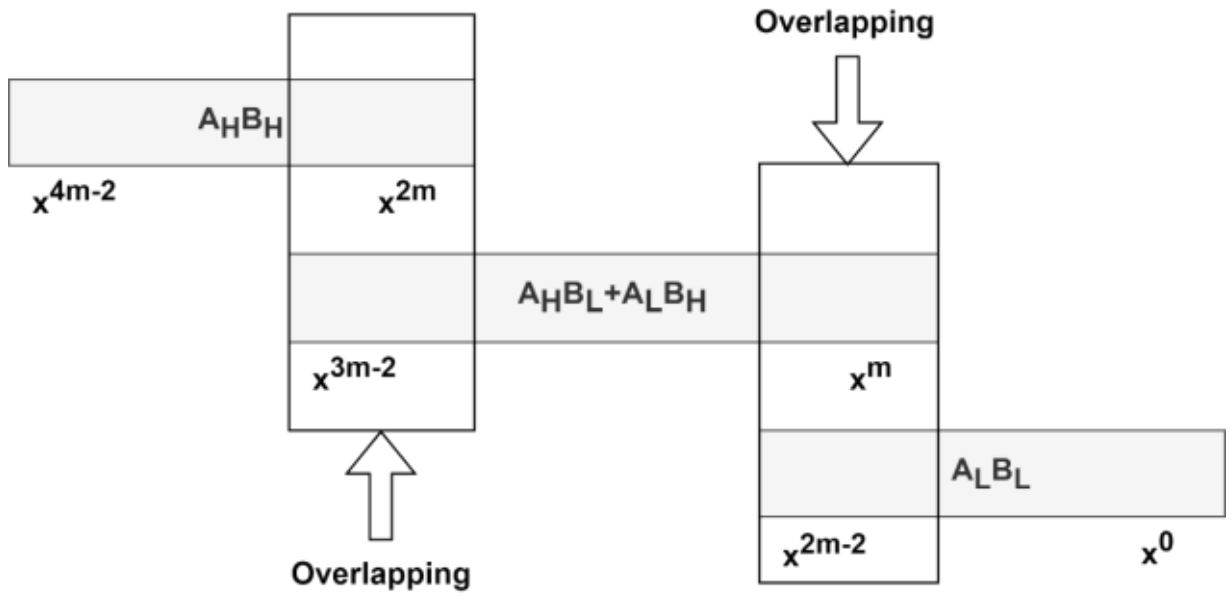

(b)

**Fig. 1.** (a) Schematic realization of $n = 2m$ bit Karatsuba Algorithm (KA) for polynomial multiplication (b) Overlap circuit.

For designing a 163-bit Karatsuba multiplier, a recursive multiplier technique is employed. For instances where the operand size is even, the operand $a(x)$ is partitioned into two equal-sized components, namely $A_H$ and $A_L$. Conversely, in cases where the operand size is odd, $A_H$ and $A_L$ do not possess equal sizes. To address this disparity, a supplementary zero is appended at the Most Significant Bit (MSB) position. This recursive process is iteratively applied to each stage of odd operand values. In this paper, we present the implementation of a 163-bit Karatsuba multiplier using the aforementioned technique as shown in Fig. 2:

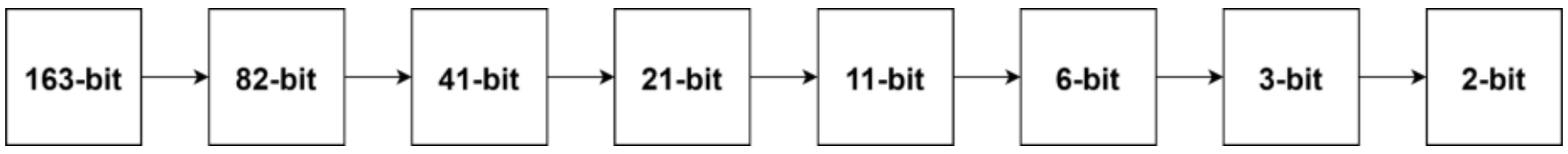

**Fig. 2.** Karatsuba multiplier for operand size 163.

Several observations can be derived from the utilization of this technique. Firstly, it is evident that the number of stages exhibits a logarithmic growth pattern relative to the size of the operand. For instance, in the case of 163-bit operand, it can be decomposed into 21-bit operand within 3 stages. However, to further decompose these 21-bit operands, an additional number of 4 stages are necessary. Consequently, the increase in number of stages leads to an increased delay. Another observation is the requirement for zero padding in multiple stages. This contributes to an increase in the complexity of the overall area required for the implementation.

The Karatsuba multiplier offers a significant advantage in terms of its low space complexities, which simplifies the implementation of multipliers for large values of $n$. However, it is important to note that in scenarios where the operand size is small, computational delay becomes a critical factor to consider. Under ideal hardware conditions, the number of gates and the delay equation for the Karatsuba multiplier are presented below [12]:

$$KM_{XOR}(n) = 6n^{log_2 3} - 8n + 2 \tag{18a}$$

$$KM_{AND}(n) = n^{log_2 3} \tag{18b}$$

$$T_{KM}(n) = T_a + (3\lceil log_2 n \rceil - 1)T_X \tag{18c}$$

where $KM_{XOR}(n)$ and $KM_{AND}(n)$ represent the number of XOR and AND gates respectively and $T_{KM}$ denotes the critical delay of the Karatsuba multiplier implementation for $n$-bit size. Here $T_a$ and $T_X$ denote the delay of an individual AND and XOR gates respectively.

It can be observed that the rate of increase in critical delay is inversely related with the operand size. To illustrate this point, the critical delay difference for 32-bit and 16-bit Karatsuba is only $3T_X$ which is same as critical delay difference for 8-bit and 4-bit Karatsuba multiplier. This makes it inefficient to implement large size Karatsuba multiplier using slow smaller-sized Karatsuba multipliers.

In this paper we have proposed a hybrid Karatsuba multiplier by utilizing the conventional Polynomial multiplier in place of an intermediary Karatsuba stage, thus requiring no further Karatsuba decompositions. This is shown in Fig. 3 where a 41-bit Polynomial multiplier is utilized to achieve this. The analysis behind the choice of 41-bits for our purpose is presented subsequently.

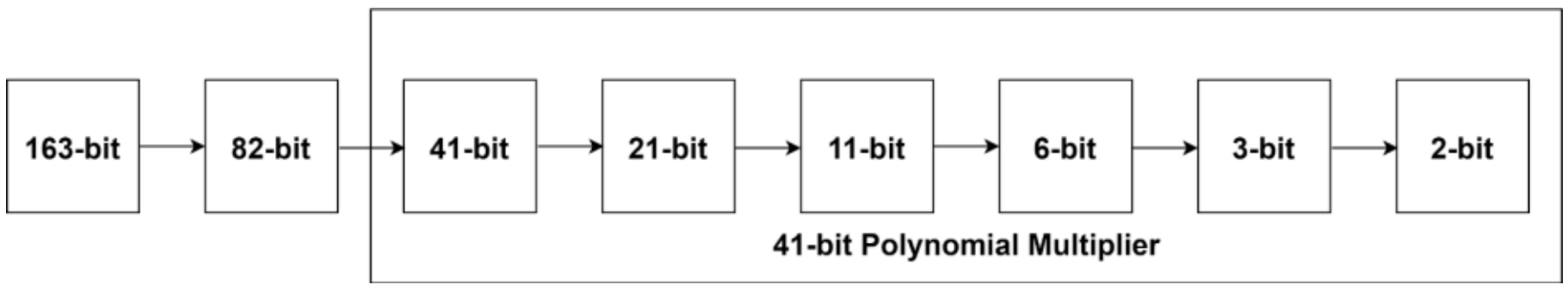

**Fig. 3.** Proposed 163-bit hybrid Karatsuba multiplier.

Assuming ideal hardware condition, the equations for number of gates and critical path delay for Polynomial multiplier are given below [11]:

$$PM_{XOR}(n) = (n-1)^2 \tag{19a}$$

$$PM_{AND}(n) = n^2 \tag{19b}$$

$$T_{PM}(n) = T_a + \lceil log_2 n \rceil T_X \tag{19c}$$

where $PM_{XOR}(n)$ and $PM_{AND}(n)$ represent the number of XOR and AND gates respectively and $T_{PM}$ denotes the critical delay of the Polynomial multiplier implementation for $n$-bit size.

A comparison between the set of equations in (18) and (19) shows that area complexity for Karatsuba is of order $n^{1.58}$ which is better than Polynomial multiplier whose area complexity is of order $n^2$. However, the critical delay for Karatsuba is three times greater than Polynomial multiplier of same size. Therefore, a tradeoff between area and delay is required in order to know at which stage of the Karatsuba multiplier a replacement with Polynomial multiplier is required for maximum performance.

If Polynomial multiplier is implemented after '$k$' stages of Karatsuba multiplier then the number of gates and delay equation for hybrid multiplier is found as below:

$$HM_{XOR}(n) = 3^k\left(\frac{n}{2^k} - 1\right)^2 + 8n\left(\left(\frac{3}{2}\right)^k - 1\right) - 2(3^k - 1) \tag{20a}$$

$$HM_{AND}(n) = 3^k \left(\frac{n}{2^k}\right)^2 \tag{20b}$$

$$T_{HM}(n) = T_a + 3kT_X + \left\lceil log_2\left(\frac{n}{2^k}\right)\right\rceil T_X \tag{20c}$$

To determine the most suitable size of the Polynomial multiplier for our proposed hybrid scheme, we conducted an analysis of LUT utilization, Delay, and Area-Delay Product (ATP) for various operand sizes comparing Karatsuba and Polynomial multipliers. Our findings are shown in Table 2 and Table 3. They indicate that the Polynomial multiplier and Karatsuba multiplier exhibit similar LUT utilization up to an operand size of 41 bits. However, as the operand size increases, the Polynomial multiplier requires significantly higher number of LUTs compared to the Karatsuba multiplier. Notably, a 163-bit Karatsuba multiplier requires approximately 22% fewer LUTs than the Polynomial multiplier. Comparison analysis of Karatsuba and Polynomial multipliers in terms of LUT, Delay and ATP as a function of operand size, implemented on the Xilinx Virtex-7 platform is given in Table 2 and Table 3.

**Table 2.** Karatsuba Multiplier Implemented on the Xilinx Virtex-7 FPGA.

| Operand Size | LUT | Delay (ns) | ATP |
|---|---|---|---|
| 6 | 16 | 6.002 | 96.03 |
| 11 | 58 | 7.059 | 409.42 |
| 21 | 206 | 9.083 | 1871.10 |
| **41** | **695** | **10.562** | **7340.59** |
| 82 | 2306 | 13.280 | 30623.68 |
| 163 | 7762 | 20.282 | 157428.88 |

**Table 3.** Polynomial Multiplier Implemented on the Xilinx Virtex-7 FPGA.

| Operand Size | LUT | Delay (ns) | ATP |
|---|---|---|---|
| 6 | 15 | 5.718 | 85.77 |
| 11 | 49 | 6.363 | 311.79 |
| 21 | 185 | 8.116 | 1501.46 |
| **41** | **694** | **9.655** | **6700.57** |
| 82 | 2599 | 12.031 | 31268.57 |
| 163 | 9982 | 18.129 | 180963.68 |

Furthermore, it is worth noting that, on average, the Polynomial multiplier outperforms the Karatsuba multiplier by around 12% in terms of delay (Table 2 and Table 3). To compare the efficiency of both the algorithms at each operand size, we calculated the area-delay product as the product of LUT utilization and delay. Since a lower value of the area-delay product indicates better performance, the algorithm with the lower ATP is considered to be superior.

Based on our analysis, we observed that the Karatsuba multiplier exhibits greater efficiency than the Polynomial multiplier for operand sizes larger than 41 bits. However, for operand sizes of 41 bit and smaller, the Polynomial multiplier proves to be more efficient. For instance, the Polynomial multiplier demonstrates around 9% lower ATP than the 41-bit Karatsuba multiplier.

Considering these factors, we have proposed utilizing a 41-bit Polynomial multiplier as the fundamental building block for constructing a 163-bit multiplier using the Karatsuba algorithm. This approach leverages the efficiency of the Polynomial multiplier for smaller operand sizes, while benefiting from the reduced LUT utilization of the Karatsuba multiplier for larger operand sizes.

# 4 Results

In our study, we have conducted the implementation of ECPM on the Xilinx Virtex-7 xc7v585tffg1761-3 device on FPGA, focusing specifically on the binary curve NIST B-163 as defined by the National Institute of Standards and Technology (NIST) [18]. The specific curve parameters for NIST B-163, essential for this implementation, are presented in Table 4 for comprehensive reference.

**Table 4.** NIST Standard Parameters over GF ($2^{163}$) [18].

| Parameter | Value |
|---|---|
| $m$ | 163 |
| $F(x)$ | $x^{163} + x^7 + x^6 + x^3 + 1$ |
| $a$ | 00000000000000000000000000000000000000001 |
| $b$ | 20a601907b8c953ca1481eb10512f78744a3205fd |
| $G$ | 3f0eba16286a2d57ea0991168d4994637e8343e36,<br>00d51fbc6c71a0094fa2cdd545b11c5c0c797324f1 |
| $n$ | 040000000000000000000292fe77e70c12a4234c33 |

The hardware computation flow for ECPM was implemented in accordance with Algorithm 2. The implementation utilized the proposed hybrid Karatsuba multiplier as the modular multiplier, accompanied by the modular squarer and inversion circuit over the finite field.

Squaring circuit was realized following the approach described in [14]. Although squaring could have been implemented using the available multiplier circuit, our paper opted for an alternative method utilizing precomputed equations and exclusively XOR gates. This approach resulted in a squaring circuit with significantly reduced critical delay and improved area utilization compared to the multiplier circuit. Both the multiplier and squaring operations were executed within a single cycle.

For the inversion circuit, the Extended Euclidean Algorithm (EEA) was employed, following the methodology as described in [13]. The implementation of the Inversion circuit requires a maximum of $2m$ cycles, where '$m$' represents the degree of the irreducible polynomial of GF($2^m$). Thus, for GF ($2^{163}$), the execution of the inversion process necessitates a maximum of 326 cycles. It is important to note that this design choice resulted in a trade-off, as the circuit is aimed for lower latency at the expense of increased area utilization.

The results of the experiments conducted in this study are presented in Table 5, which compare the area utilization, frequency of operation, clock cycle time, total computation time, and the area-time product of the proposed design as well as three referenced works [19-21]. In [19], the authors have employed a bit-serial multiplier, thereby, optimizing LUT usage at the cost of increased execution cycles. Further, in [20], the authors have segregated adder, squarer, and multiplier into separate cycles, neglecting hardware parallelism, during the implementation of point multiplication. Where as in [21], a 3-bit parallel multiplier has been utilized for point multiplication, thus, curtailing cycles but inflating LUT consumption, culminating in a higher area-time product.

**Table 5.** Comparison of Proposed Method with Previous Implementations of ECPM ($Q$=$kP$) over GF ($2^{163}$).

| Design | Area (LUTs) | Freq. (MHz) | Clock Cycles | Time (μs) | Area-Time Product (× 1000) |
|---|---|---|---|---|---|
| Nyugen [19] | 3806 | 800 | 52012 | 65.0 | 247.39 |
| Imran [20] | 10128 | 135 | 3426 | 25.4 | 257.25 |
| Khan [21] | 41090 | 159 | 450 | 2.83 | 116.28 |
| **This Work** | 14195 | 213 | 1298 | 6.09 | 86.45 |

Our proposed design was realized using 14195 LUTs which is relatively higher in comparison to [19] and [20]. However, it achieved a significantly higher operating frequency of 213 MHz, with respect to [20] and [21]. Consequently, the total computation time was reduced to 6.09 microseconds, showcasing the improved efficiency (in terms of lower ATP) of our design, compared to [19], [20], and [21].

Moreover, when considering the area-time product which is a parameter of overall efficiency, our proposed design achieved a substantially lower ATP value of 86.45 x1000 LUT-$\mu$s, indicating a favorable balance between area utilization and computation time. Further, the proposed ECPM design was subjected to power analysis, providing an estimate of 0.98 watts of power dissipation. These results demonstrate the effectiveness and competitiveness of our design in terms of performance and efficiency compared to the referenced works.

## 5 Conclusion

In this paper a hybrid Karatsuba multiplier has been implemented that utilizes small bit-sized conventional Polynomial multipliers. It provided an improved solution for larger bit sized modular multipliers by effectively balancing area efficiency and critical path delay. The hybrid Karatsuba multiplier is integrated with a finite field adder, squarer, and Extended Euclidean inversion circuit to implement Elliptic Curve Point Multiplication (ECPM) utilizing projective coordinate-based Montgomery algorithm. The architecture demonstrated superior performance for ECPM computation on the Xilinx Virtex-7 FPGA platform. The evaluation results showcased optimized trade-off between area utilization and computation time, resulting in enhanced efficiency compared to existing architectures. Overall, the proposed hybrid Karatsuba multiplier offers a promising and efficient solution for ECPM computation.